\documentclass{article}

\usepackage[letterpaper,margin=1in]{geometry}
\usepackage[latin1]{inputenc}
\usepackage[T1]{fontenc}
\usepackage[english]{babel}
\usepackage{amsmath,amsthm,amssymb,amsfonts}
\usepackage{url}

\newtheorem{theorem}{\large Theorem} 

\newtheorem{remark}[theorem]{\large Remark}

\font\tenmath=msbm10

\newfam\mathfam \textfont\mathfam=\tenmath

\def \\ { \cr }

\newcommand{\J}{{\cal J}}

\newcommand{\R}{{\cal R}}

\newcommand{\uR}{{\underline{\R}}}
\newcommand{\uJ}{\underline{\J}}
\newcommand{\uj}{\underline{J}}
\newcommand\Rcs{\R^{\mbox{cs}}}
\newcommand\RRcs{R^{\mbox{cs}}}
\newcommand\norm[1]{\Vert#1\Vert}

\begin{document}

\title{Reproduction matrix for an epidemic and lockdowns in a city}
\author{Vicente Acu\~na \and  Mar\'ia Paz Cort\'es \and Andrew Hart \and Servet
Mart{\'\i}nez \and Juan Carlos Maureira
\\ \\
\parbox{5in}{\rmfamily\upshape\normalsize
Center for Mathematical Modeling, 
UMI 2071 CNRS-UCHILE, 
Facultad de Ciencias F\'isicas y Matem\'aticas,
Universidad de Chile, 
Santiago, Chile.}}
\maketitle

\begin{abstract}
We consider an epidemic spreading in a city which 
is divided geographically into
different districts. We introduce the reproduction matrix
$\R=\bigl(R(i,j)\bigr)$ between districts, where $R(i,j)$ is the mean
number of individuals in district $j$ infected by an individual from district $i$. 
We analyse policies of
partial lockdowns of the city, that is of a set of districts, based on the
study of matrix $\R$, where rows and columns corresponding to districts in
lockdown are set to zero.  This schema can also be applied to a country divided
into regions or other appropriate units, provided the relevant 
information is available.
We conclude by analyzing a matrix~$\R$ which was constructed for the spread of
COVID-19 in Santiago, Chile, with the aid of an agent-based simulator for generating surrogate district data.
\end{abstract}

\bigskip

\section{Introduction}
\label{sec:intro}

\noindent We are interested in studying the spread of an epidemic 
in a city divided into districts. The city is assumed to be isolated. 
We require the population to have attained some stable 
equilibrium in terms of social behavior.  
We shall assume that the only policy decision that the city authority has 
at its disposal for confronting the epidemic is to impose
a lockdown on a set of districts during some period of time. In
practice, one would expect this set of districts to vary throughout the
evolution of the epidemic.

\medskip

In addition, we assume that during a lockdown, city
inhabitants will modify their social behavior ideally in order to comply with
the lockdown. Nevertheless, based on a number of partial lockdowns instituted by
the government in Santiago, the capital of Chile, as well as the level of
economic activity and daily commuting by public and private transport that
typically prevails in this city, it is not certain that the population will achieve the necessary degree of
compliance for a lockdown to be effective.

\medskip

The motivation for this work is in reaction to the COVID-19 epidemic
currently being suffered in Santiago, where
the inhabitants throughout the city use masks and practise social distancing
measures, and where schools, universities and other education facilities 
were closed and the teaching of classes moved on line where possible  near the
commencement of the outbreak. On the other hand, the Chilean authorities have decreed a
series of partial dynamic lockdowns of various districts in Santiago. 
Our intent here is to try and provide a more theoretical  basis for these kinds
of decisions. 

\medskip

In the next section, we introduce the reproduction matrix $\R$ and explain the 
key role of its Perron-Frobenius eigenvalue for describing the behavior
of~$\R$ as it is iteratively applied to the evolution equation.
This eigenvalue essentially constitutes the effective reproduction 
number which
determines if the epidemic explodes, persists or becomes extinct.

\medskip

In Section~\ref{sec:evolution},  we describe the spread of the epidemic
in terms of the evolution of powers of the reproduction matrix. In order to do
this, we make two simplifications of the epidemic evolution:
\begin{enumerate}
  \item 
there exists some period of time such that once an individual is infected
at, say,  time~$t$, he/she will only be contagious during the next period of
time at $t+1$; and 
\item
the size of the population of susceptible
individuals is on the order of the original population size.
\end{enumerate}
The last assumption is realistic for Santiago because it is currently 
in the initial exponential-growth phase of the epidemic. The first
assumption is less realistic as the period of time corresponds to the mean
generational time-to-infection, which has been
estimated to be around~$4$ or~$5$ days for COVID-19 \cite{Nishiura&etal2020}.

\medskip

In Section~\ref{sec:lockdown}, we describe how lockdowns or sanitary
cordons modify the structure of the matrix and the Perron-Frobenius eigenvalue.

\medskip

The data necessary for estimating the matrix~$\R$ is similar to the kind of
data collected during contact tracing. However, contact tracing is labor
intensive and more automated ways of collecting this kind of data, such as
via smart phone apps and Bluetooth-enabled devices, come with a host of
issues relating to privacy and use/misuse which many governments are currently
wrestling with.
At present, such data is generally not available, so we have elected to use
a strategy whereby we obtain surrogate data by means of simulation in order to
illustrate how~$\R$ can be used to decide which districts to include in a
lockdown. Section~\ref{sec:simulation} describes the agent-based simulation model
used to generate the surrogate data and obtain an estimation of the reproduction
matrix $\R$ for the city of Santiago. In this model, agents represent
people who can become infected by interacting with infected agents.
They are placed in a grid so as to reproduce the real limits
and densities of the city (and its districts) and to emulate daily commuter
traffic-flow patterns based on real public transportation data.
We explain how this model can be used to describe the propagation of the
infection through the city and between the districts, and how the matrix $\R$
is estimated from a set of simulation replicates.
We note that density, transportation and the spatial configuration of
districts are the main characteristics determining the variability in the
values of elements of the matrix~$\R$.  

\medskip

Finally, the conclusions presented in Section 6 focus on the lockdowns that Santiago currently faces. In particular, by computing the spectral radius of
the matrix~$\R$, we can estimate the reproduction number for different
scenarios:
(a) When there is no district in lockdown; (b) when there is only a single
district in lockdown (computed for each district under consideration), and  (c)
when there is a set of districts in lockdown.

\section{The reproduction matrix}
\label{sec:matrix}

\noindent Let $I_m=\{1,2,\ldots,m\}$ be the set of districts of a city, 
the population size of the city is denoted by $N$ while $N_i$ denotes the
population of district $i$. 

\medskip

\noindent Define the matrix $\R=(R(i,j): i,j=1, \ldots, m)$ as follows: 
$R(i,j)$ is the mean number of individuals residing in 
district $j$ that are infected by an individual from district $i$ while
the individual is contagious. 

\medskip

\noindent Hence $R_i=\sum_{j=1}^m R(i,j)$ is the overall mean 
number of people infected by an individual from district $i$, irrespective of
the district they reside in.
If $R_i$ is the same for all districts, that is, $R_i=R$ for
$i=1,2,\ldots,m$, then $R$ is equivalent to the effective reproduction number $R$.

\medskip

\noindent We note here that in order to have
$R(i,j)>0$ there must be contact between the populations of $i$ and $j$, in 
which case one is highly likely to have $R(j,i)>0$. 
There is no loss of generality in assuming the matrix $\R=(R(i,j): 
i,j=1,\ldots,m)$ is irreducible,
which means that for any pair of districts $i,j$, there exists a path
of infections (most likely passing through a number of districts) from an
individual in $i$ to some individual in $j$, that is $\R^t(i,j)>0$ for some
power $t>0$. When there is no irreducibility, the components can be studied
separately, but irreducibility is a very weak condition to assume in a city.

\medskip

From the Perron-Frobenius theory \cite[Chapter~1]{es81}, there exists an
eigenvalue $\rho(\R)>0$ that satisfies the following conditions: It has unique (up
to a constant multiple), strictly positive left and right eigenvectors
$\ell=(\ell(i): i\in I_m)$ and $h=(h(i): i\in I_m)$ satisfying
$$
\ell' \R=\rho(\R) \ell' \mbox{ and } \R h=\rho(\R) h.
$$
Here the vectors are column vectors and $'$ signifies the transpose so
that $\ell'$ is a row vector. Note that the vectors $\ell$ and $h$ can be
normalized such that $\sum_{i=1}^m \ell(i) h(i)=1$. With this normalization,
the matrix with elements
$$
\rho^{-1}R(i,j)h(j)/h(i), \quad i,j\in I_m,
$$
is stochastic with stationary distribution
$(\ell(i)h(i): i\in I_m)$. In the following discussion, we shall assume that the
eigenvectors have been normalized in this way.

\medskip

\noindent $\rho(\R)$ is simple and it is the spectral radius of $\R$, so
it satisfies 
$$
\rho(\R)=\lim\limits_{t\to \infty} {\norm{R^t}}^{\frac{1}{t}},
$$
for any norm $\norm{\cdot}$.
It gives the geometric rate of the asymptotic behavior of 
iterative multiplication by $\R$, that is when it is successively 
multiplied by itself (powered up). One can specify the 
precise limit behavior of iterating
the matrix $\R$: 
$$
\R^t=\rho(\R)^t h \ell' + o\bigl((\rho(\R))^t\bigr),
$$
where $o\bigl((\rho(\R))^t\bigr)$ is a matrix satisfying $\rho(\R)^t
o\bigl((\rho(\R))^t\bigr)\to 0$ as $t\to \infty$. 

\medskip

\noindent The matrix $A=h \ell'$, where $A(i,j)=h(i)\ell(j)$,
is strictly positive for $i,j=1,\ldots,m$. Therefore, as $t$ becomes
large, $\R^t$ has the following limit behavior:
$$
\lim_{t\to\infty} R^t(i,j)=
\begin{cases}
\infty, & \mbox{ if } \rho(\R)>1,\\
h(i)\ell(j), & \mbox{ if } \rho(\R)=1, \\
0 & \mbox{ if } \rho(\R)<1.
\end{cases}
$$
In particular one has
\begin{equation}
\label{limit1}
\lim\limits_{t\to \infty} \R^t=0 \;\Leftrightarrow\; \rho(\R)<1.
\end{equation}

\begin{remark}
\label{extension}
If $R_i$ is the same for all districts, $R_i=R$ for 
$i=1,2,\ldots,m$,
then $\rho(\R)=R$ is the effective reproduction number $R$. In fact, if $R=1$,
the matrix $\R$ must be stochastic and then $\rho(\R)=1$ and the other cases
follow easily (it suffices to argue that an iteration of a stochastic 
matrix is stochastic). Hence, our 
study extends the notion of an effective reproduction number to the
introduction of an effective reproduction matrix $\R$. 
\end{remark}

\medskip

\begin{remark}
\label{rem:extcases}
Note that one can have
$\rho(\R)<1$ even if there exist some districts with $R_i>1$ because 
the set of districts with $R_i<1$ may compensate 
enough to obtain a spectral radius strictly smaller than one, ensuring the
extinction of the epidemic.
See Table~\ref{tab:3} for an example of this. 

We note that the assumption of irreducibility implies that
if all but one of the $R_i$'s are constant, that is, for some $i_0$, $R_i=r$ for
$i \neq i_0$, then one can check that $\rho(\R)$ is less than, equal to 
or
greater than~$r$ according as $R_{i_0}$ is less than, equal to or greater
than~$r$ respectively.
\end{remark}

\section{Evolution of total infected individuals}
\label{sec:evolution}
 
The matrix $\R$ captures the complex interaction between the different
districts and the resulting effect on the spread of the disease. Due to this,
it will be our primary tool for studying the evolution of the epidemic.
We make the strong assumption that any infected individual will infect other
individuals after some unitary period of time. That is, if individual~$i$
becomes infected at time~$t$ and~$i$ infects individual~$j$, then~$j$ 
becomes infected exactly at time $t+1$. 
The unitary period of time could represent a fixed number of days, 
typically the mean time between a primary and secondary 
infection. For instance, this time is reported to be around~$4$ 
or~$5$ days for COVID-19 \cite{Nishiura&etal2020}. 

\medskip

Let $J_i(s)$ be the total number of 
infected individuals at time $s$ in district $i$ and let 
$\J(s)=\bigl(J_i(s): i\in I_m\bigr)$ be the vector of infected people
in the city at time $s$. For stating the evolution 
equation, we assume we are in the initial growth phase of the epidemic so that 
the number of susceptible individuals in district $i$ will be 
on the order of $N_i$. Then, the evolution equation is
$$
\J'(s+1)=\J'(s) \R, \mbox{ that is, } 
J_j(s+1)=\sum_{i=1}^m J_i(s) R(i,j), \quad s=0,1,\ldots.
$$ 
Iterating the equation yields
\begin{equation}
\label{evolution}
\J'(s+t)=\J'(s) \R^t, \quad t=1,2, \ldots \mbox{ and } s=0,1,2,\ldots.
\end{equation}

\medskip

Now, from (\ref{limit1}) and~(\ref{evolution}), one has 
$\lim\limits_{t\to \infty}
\J(t)=0$ if and only if the condition $\rho(\R)<1$ holds. Consequently, 
the epidemic dies out if $\rho(\R)<1$. 
In contrast, the infected population explodes
geometrically when $\rho(\R)>1$ while if
$\rho(\R)=1$, the epidemic continues to spread steadily at a constant rate. 
In the latter two cases, the viral epidemic eventually 
infects the entire population since
the population size is finite. 

\medskip

\begin{remark}
Let $d_i$ be
the probability that an infected individual from district $i$ dies from the 
illness, so that with probability $1-d_i$ an infected 
individual recovers. Further, since the population is finite, the epidemic is
bound to die out. The spectral radius~$\rho(\R)$ merely determines the manner in
which it  spreads before the extinction time~$T$ has lapsed.

\begin{itemize}
\item For $\rho(\R)>1$, irreducibility implies that the whole population of the
city becomes infected geometrically fast (in time proportional to the logarithm
of the population size~$N$).
Consequently, the number of people that die in district $i$ is expected to  
be $d_i N_i$ and the total number of people that are
expected to die during the course of the epidemic is $\sum_{i=1}^m d_i N_i$.

\item If $\rho(\R)=1$, the epidemic spreads at a constant rate,
infecting the entire population in a time~$T$ proportional to the population size~$N$.
However, the total number of deaths expected to occur by the time the 
epidemic has run its course is
once again $\sum_{i=1}^m d_i N_i$, the same as when $\rho(\R)>1$.

\item Finally, when $\rho(\R)<1$ the epidemic becomes extinct geometrically
fast.
The total number of infected people in district~$i$ is $\sum_{s=0}^T J_i(s)$, 
and the expected number of deaths for the district is $d(i)
\sum\limits_{s=0}^T \J_i(s)$. Thus, the total number of infected cases in the 
population during the epidemic is $\sum\limits_{i=1}^m \sum\limits_{s=0}^T 
J_i(s)$ and
the total number of deaths is expected to be $\sum\limits_{i=1}^m d(i)
\sum\limits_{s=0}^T \J_i(s)$.
\end{itemize}
\end{remark}

\section{ Lockdown }
\label{sec:lockdown}

When the authorities declare a lockdown for some districts, 
transmission of the infection between those districts and the districts
not in lockdown theoretically halts.
Community transmission continues to take place within the districts under
lockdown, for instance, transmission among people sharing the same living space
during the lockdown increases, but theoretically it should cease between
people in the same district who do not live together. Infections
in these locked-down districts is not due to the epidemic spreading within
the district but caused by healthy and infected
cohabitants being confined to shared (but disjoint)  living spaces, so we will
consider this separately.

\medskip

Let $L\subset I_m$ be the set of districts that are put into lockdown. 
Then, the new matrix of the mean number of infections per contagious
individual $\R_L=\bigl(R_L(i,j): i,j\in I_m\bigr)$ becomes
$$
R_L(i,j)=
\begin{cases}
R(i,j) & \mbox{ if } i,j\notin L, \\
0 & \hbox{ if } i\in L\mbox{ or } j\in L.
\end{cases}
$$
This is no more than the truncation of~$\R$ to
the set of districts~$L^c$, the complement of~$L$.
Since there are no interactions between the districts in $L$ and~$L^c$, the
infection can only spread among districts in $L^c$ which are not subject to lockdown 
(as viral spread is essentially restricted  to the shared 
living spaces of districts in $L$). So it
is convenient to restrict the study of the evolution of the population to 
districts in $L^c$.

\medskip

Note that the restriction of $\R$ and $\R_L$ to the states in $L^c$
is the same. We set
$\uR=\bigl(R(i,j): i,j\in L^c\bigr)$ to be the matrix of mean
infections for the districts that are not in lockdown. 
Let $\uJ(s)=\bigl({\uj}_i(s): i\in L^c \bigr)$ 
be the evolution of the infected individuals in districts
in $L^c$. Note that the truncation is imposed from the moment the lockdown of
$L$ is decreed and so the initial point ${\uj}_i(0)$ is the number of 
infected people
in district $i\notin L$ when the lockdown starts. The evolution of the
infected population vector during the lockdown is given by 
$$
{\uJ}'(s+t)={\uJ}'(s)\uR^t.
$$
We note that the spectrum of $\uR$ is the same as that for $\R_L$ 
and so the common spectral radius of $\uR$ and $\R_L$ will be denoted by
$\rho(\R_L)$.

\medskip

It is possible that the lockdown framework splits $\uR$ into a number of 
irreducible components because the infection from district $i$ to $i'$, both in 
$L^c$ might only be due to agents in districts in $L$.
When $\uR$ is irreducible, the analysis
carried out on~$\R$ remains valid on~$\uR$. If not one must analyse the 
distinct irreducible components and the spectral radius $\rho(\R_L)$ of
$\uR$ is the maximum of the spectral radii of the restrictions 
of $\uR$ to these components. 

\medskip

The choice of the set of districts~$L$ needs to ensure that the
largest spectral radius $\rho(\R_L)$ is strictly smaller than~$1$ in order to 
bring the epidemic under control.
Note that the minimum value of $\rho(\R_L)=0$ is only obtained at $L=I_m$ when
the whole city is in lockdown. 
However, choosing~$L$ to be the set of districts giving the biggest
$\rho(\R_L)$ strictly smaller than $1$ may not be the best strategy because
$\rho(\R_L)$ may be too close to~$1$ to end the epidemic due to random fluctuations in
incidence as well as uncertainty in the data used to compute
$\rho(\R_L)$.
The key point is that in reality, there are real-world constraints that need to
be considered (such as the mean time that people are in lockdown) and this leads
to a multi-objective optimization problem.

\medskip

\begin{remark}
Since we are choosing the districts $L$ that are 
to be locked down so as to 
obtain a spectral radius of $\rho(\R_L)<1$, it is the asymptotic behavior of
the matrix $\R_L$ that will govern the evolution of the system during the
period the lockdown is in force.
Thus, the total number of infected people will 
decrease geometrically. In district $j\notin L$, the total number of 
infections that result will be $\sum_s \uj_j(s)$ and so the final number
of infections in the districts not under lockdown 
will be $\sum_{j\notin L}\sum_s \uj_j(s)$. 
To obtain the total number of infections in the
population from the time of imposing the lockdown
it is necessary to add the quantity $\sum_{i\in L} \gamma_i M_i$, where 
$M_i$ is the total number of healthy people cohabiting with 
infected individuals in district ~$i$ during the lockdown and $\gamma_i$ is
the probability that a susceptible individual cohabiting with an infected
individual becomes infected during the lockdown.
The number of deaths expected to occur during the imposition of the total
lockdown, provided the lockdown is in force long enough to bring the epidemic to
extinction (or nearly to extinction) is:
$$
\sum_{j\notin L} d_j\sum_s \uj_j(s) + \sum_{j\in L} d_j \gamma_j M_j.
$$
\end{remark}

\begin{remark}
The policy of `cordon sanitaire' or `sanitary cordon' consists of disconnecting
two geographic regions by preventing movement between them. In a city, this means
selecting two groups of districts, both of which are internally connected so that
individuals may move freely from district to district within each group, while
the connections bridging them are suppressed.
This is a weaker measure than the lockdown discussed above. If $L$ and
$L^c$ are the two groups, the new matrix~$\RRcs$ encapsulating the
cordon sanitaire policy must preserve the values $R(i,j)$ for districts~$i$
and~$j$ inside the same group, but set $\Rcs(i,j)=0$ when districts~$i$ and~$j$
belong to~$L$ and~$L^c$ or to $L^c$ and $L$ respectively. Thus, 
$\RRcs=\bigl(\Rcs(i,j): i,j\in
I_m\bigr)$, where
$$
\Rcs(i,j) = \begin{cases}
R(i,j), \mbox{if } i,j\in L \mbox{ or } i,j\in L^c, \\
0, \mbox{otherwise.}
\end{cases}
$$
The spectral radius $\rho(\RRcs)$ of this
block diagonal matrix, which is given by the maximum of the spectral radii of
the blocks, will be strictly smaller than $\rho(\R)$ in general and in order for the cordon sanitaire policy to be effective, it must be strictly smaller than~$1$.
\end{remark}\medskip

\section{The simulator}
\label{sec:simulation}

To evaluate the effectiveness and applicability of the 
simple model proposed here, we computed
a reproduction matrix $R(i,j)$ for the city of Santiago in Chile, at about one
month into the outbreak  by using the data obtained from a simulation of the
epidemic's evolution  in the city. As data detailing the chain of infection of
individuals, together with their districts of residence was not available,
it was necessary to obtain surrogate data to use in its stead.

\medskip

We have developed an agent-based simulator which is capable of generating the
chain of infection of each individual for a city. 
Within this simulator, the city of Santiago was
represented considering the 40 most important districts of the greater 
Santiago area (part of the metropolitan 
region) with their corresponding population densities \cite{censo2017} and 
commuting flows \cite{dtpm}. 
A brief description of the simulator follows.

\medskip

The simulator consists of a collection of agents representing a population of
individuals that interact with each other following a set of rules:

\begin{itemize}
\item At each simulation step (representing a day), 
each agent  has one of the following states: 
Susceptible, Latent (incubating the virus),
Asymptomatic and contagious, Symptomatic and contagious, Hospitalized in a
normal bed, Hospitalized in an ICU bed,  Hospitalized in a normal bed but
waiting for a ICU bed (there is a finite number of them), Recovered and
Deceased.
 \item The area (or playground) inhabited by the agents and where they interact 
 is modeled as a grid of cells onto which a map of Santiago is projected,
 together with the boundaries demarcating the districts.

\item The position and age of each agent is 
assigned according to population densities and age distributions reported by
district in census records \cite{censo2017}.

\item At each simulation step (representing a day), each agent 
moves from its starting position (home) to another cell in a designated
destination district (representing work/school) according to a
commuting matrix\cite{dtpm} indicating the flows between districts 
\footnote{The commuting matrix was obtained from the public transportation
operator Transantiago and is current as at 10 September, 2019.}. 
Then, each agent returns to their home cell where 
they have the opportunity to interact with neighbors in their vicinity.

\item Each agent interacts with other nearby agents according to an interaction
radius. A susceptible  agent is infected by a contagious agent
in their vicinity according to some fixed probability.
\end{itemize}

Agent states evolve day by day in response to interaction with other agents
throughout the simulation timeline (where time is represented as a series of
simulation steps).
The agents' level of daily interaction is modeled by adjusting the size of their
neighborhoods, their travel distance, and the probability of contact with
neighboring agents.
Recording the state transitions made by agents over the simulation time allows
the evolution of the epidemic to be tracked and analyzed by means of different
daily indicators such as incidence, prevalence, mortality, ICU occupancy.

\medskip

In addition, the simulator implements a set of template rules that enable the
behavior of agents to be changed at any step of the simulation according to
various selection criteria such as the interaction and infection probabilities,
mobility ranges (governing travel distances) and neighborhood sizes (which
determine how close agents must be in order to interact).
In this way, mitigation policies imposed by the health authority such
as curfews, social isolation, total quarantine, school closures and isolation of
the elderly or very young 
can be emulated directly as part of the simulation scenario.

\medskip

This simulator integrates concepts from the simulation models used
by Ferguson et al.  \cite{ferguson2020report} and Donsimoni et
al. \cite{Donsimoni2020}. From the latter, we borrowed the basic idea of each
agent interacting with different groups of individuals (home, work, school). The main difference is we restrict agents to move within a target
community with a specified probability and randomly otherwise. Agents in
Ferguson's model always interact within the same community. Similar
to~\cite{ferguson2020report}, we have modeled the evolution  of the illness
within an agent as a Markov chain, defining the series of states and state
transitions that each agent must make according to specified probabilities
before eventually either recovering or dying.

\medskip

The calibration of the scenario we simulated of Santiago was done in a
two-fold procedure. On one hand, parameters were selected that allow the
reproduction of epidemic and demographic 
characteristics such as the basic reproduction number $R_0$, 
the outbreak generation time, and the density and age distribution of the
population, while on the other hand, interaction and infection rates were chosen
in order to follow the progression of new symptomatic cases reported daily
by the Chilean health authorities.

\medskip

With the simulation appropriately fitted to the outbreak under study, the
matrix~$\R$ can be estimated by simulating the desired scenario multiple times
using different random seeds to initialize the simulator in order to mimic the
stochastic process governing the evolution of the epidemic.
Note that matrix~$\R$  depends on the number of susceptibles that each
agent encounters during its contagious period, which varies during the course
of the epidemic (at the beginning, this should generally be greater than it
will be near the end).

\medskip

Therefore, to obtain a matrix that describes the effective $\R$ at a given
time (analogous to the well-known effective reproduction number~$R$), we should
consider only computing the number of infections transmitted by agents infected
within a small interval prior to that time.
Thus, $\R$ is estimated for a time~$t$ by considering the set of agents that
contract the disease in a small window prior to and including~$t$ and
calculating  the mean number of infections they caused in each district
during a single simulation execution (or simulation replicate). The set of
matrices obtained from a number of simulation replicates is then averaged
(elementwise) to provide an estimate of the reproduction matrix $\R$ for the
city.

\section{Application to Santiago}
\label{sec:application}

Simulations of a $1/9$-scaled version of 40 districts within the Greater
Santiago area were carried out using the simulator described in the
preceding section. Parameters were calibrated to enable the simulator to mimic
the overall  trend in Chile during the first 30 days of the evolution of the
epidemic (as reported daily by the Chilean health authority).
The simulation was replicated 50 times and each such evolution of the epidemic
was analyzed independently (in the time interval from day $20$ through to day
$30$) in order to compute its reproduction matrix as at day $30$.
As described in the previous section, the 50 matrices were then averaged
elementwise to obtain an estimate of the reproduction matrix $\R$ for Greater Santiago.
Being $40\times40$, the matrix is too large to include here, but
Table~\ref{tab:1} displays summary statistics for each of its
$40$ rows, together with the row sums which yield the local
reproduction number $R_i$ for each district.
The spectral radius of this matrix is $1.62$.

\medskip

Having obtained a spectral radius greater than one, we explored what happens if
a single district is placed under lockdown. We computed the spectral radius
of the matrix resulting from placing each of the $40$ districts singly
under lockdown. As described above, this is achieved by removing the row and column
corresponding to the district in lockdown from~$\R$. The spectral radii that
result are displayed in the last column of Table~\ref{tab:1}. The districts
that produce the greatest reduction in the spectral radius when in lockdown are
Puente Alto and Pirque (near $1.56$), followed by La Florida, La Pintana,
Santiago, La Cisterna and El Bosque (all with spectral radii less than $1.60$).

\medskip

Next, we examined the accumulative affect of applying lockdowns successively.
As in the previous paragraph, we examined the effect of putting a single
district in lockdown and kept the one that provided the greatest
reduction in the spectral radius. This was Puente Alto. Then, we
considered
all combinations of putting Puente Alto and a second
district under lockdown together. We kept the
combination that produced the smallest spectral radius, which was Puente Alto
and Santiago. We then added a third district to the lockdown and continued
adding districts until we 
obtained a spectral radius smaller than~$1$. Table~\ref{tab:2} summarizes 
the results of this
procedure. In the end, it was necessary to put eight districts under
lockdown: Puente Alto, Santiago, El Bosque, Pirque, Maip\'u, Cerro Navia,
La Cisterna and \~Nu\~noa. The spectral radius resulting from locking down
these eight districts is $0.984$.
The matrix~$\R$ corresponding to the districts that are not in lockdown is
$32\times32$. Its row sums are shown in Table~\ref{tab:3}, where it can be seen
that some are greater than~$1$ even though the
spectral radius is less than~$1$ (see Remark~\ref{rem:extcases}).

\begin{table}
\caption{Summary statistics for rows of the reproduction matrix estimated
for Santiago. The `lockdown' column shows the spectral radius that results when
the corresponding district is placed singly under lockdown.}
\label{tab:1}

\centering
\begin{tabular}{l|rr|rrrrr|rr}
\hline
District & Mean & Std & \multicolumn5c{Quantiles} & Row & Lock- \\
&   &    &    Min &    25\% &    50\% &    75\% &    Max & sum &  down \\
\hline
CERRILLOS           & 0.031 & 0.064 & 0.000 & 0.004 & 0.012 & 0.020 & 0.313 &    0.451 &       1.615 \\
CERRO NAVIA         & 0.042 & 0.103 & 0.000 & 0.002 & 0.009 & 0.021 & 0.525 &    1.525 &       1.609 \\
COLINA              & 0.014 & 0.037 & 0.000 & 0.000 & 0.000 & 0.007 & 0.198 &    0.387 &       1.617 \\
CONCHAL\'I            & 0.036 & 0.087 & 0.000 & 0.001 & 0.010 & 0.019 & 0.486 &    
1.095 &       1.613 \\
EL BOSQUE           & 0.043 & 0.110 & 0.000 & 0.000 & 0.004 & 0.022 & 0.576 &    1.878 &       1.599 \\
ESTACI\'ON CENTRAL    & 0.039 & 0.088 & 0.000 & 0.005 & 0.013 & 0.036 & 0.433 &    
1.291 &       1.609 \\
HUECHURABA          & 0.029 & 0.068 & 0.000 & 0.001 & 0.006 & 0.021 & 0.349 &    0.739 &       1.616 \\
INDEPENDENCIA       & 0.042 & 0.104 & 0.000 & 0.003 & 0.013 & 0.035 & 0.565 &    1.351 &       1.607 \\
LA CISTERNA         & 0.045 & 0.094 & 0.000 & 0.001 & 0.006 & 0.029 & 0.387 &    1.628 &       1.598 \\
LA FLORIDA          & 0.031 & 0.121 & 0.000 & 0.001 & 0.003 & 0.011 & 0.757 &    1.989 &       1.595 \\
LA GRANJA           & 0.040 & 0.088 & 0.000 & 0.004 & 0.015 & 0.027 & 0.493 &    1.195 &       1.604 \\
LA PINTANA          & 0.035 & 0.085 & 0.000 & 0.001 & 0.003 & 0.025 & 0.485 &    1.445 &       1.595 \\
LA REINA            & 0.025 & 0.055 & 0.000 & 0.001 & 0.004 & 0.010 & 0.236 &    0.442 &       1.616 \\
LAMPA               & 0.016 & 0.039 & 0.000 & 0.000 & 0.000 & 0.016 & 0.204 &    0.441 &       1.617 \\
LAS CONDES          & 0.024 & 0.074 & 0.000 & 0.001 & 0.003 & 0.010 & 0.458 &    1.289 &       1.616 \\
LO BARNECHEA        & 0.016 & 0.056 & 0.000 & 0.000 & 0.000 & 0.004 & 0.313 &    0.317 &       1.617 \\
LO ESPEJO           & 0.043 & 0.077 & 0.000 & 0.005 & 0.016 & 0.040 & 0.352 &    1.301 &       1.604 \\
LO PRADO            & 0.041 & 0.100 & 0.000 & 0.002 & 0.005 & 0.010 & 0.510 &    1.075 &       1.612 \\
MACUL               & 0.029 & 0.064 & 0.000 & 0.002 & 0.006 & 0.014 & 0.324 &    0.766 &       1.614 \\
MAIP\'U               & 0.035 & 0.143 & 0.000 & 0.001 & 0.005 & 0.011 & 0.905 &    
2.798 &       1.610 \\
PADRE HURTADO       & 0.023 & 0.063 & 0.000 & 0.000 & 0.001 & 0.009 & 0.314 &    0.424 &       1.616 \\
PEDRO AGUIRRE CERDA & 0.038 & 0.085 & 0.000 & 0.006 & 0.014 & 0.030 & 0.408 &    0.977 &       1.610 \\
PE\~NAFLOR            & 0.024 & 0.092 & 0.000 & 0.000 & 0.000 & 0.002 & 0.424 &    
0.564 &       1.616 \\
PE\~NALOL\'EN           & 0.031 & 0.088 & 0.000 & 0.003 & 0.005 & 0.013 & 0.518 &    
1.718 &       1.612 \\
PIRQUE              & 0.065 & 0.247 & 0.000 & 0.000 & 0.001 & 0.006 & 1.300 &    1.465 &       1.568 \\
PROVIDENCIA         & 0.034 & 0.073 & 0.000 & 0.002 & 0.013 & 0.023 & 0.377 &    1.361 &       1.612 \\
PUDAHUEL            & 0.036 & 0.099 & 0.000 & 0.002 & 0.006 & 0.015 & 0.576 &    1.462 &       1.611 \\
PUENTE ALTO         & 0.041 & 0.220 & 0.000 & 0.000 & 0.000 & 0.005 & 1.393 &    3.361 &       1.559 \\
QUILICURA           & 0.031 & 0.115 & 0.000 & 0.001 & 0.003 & 0.013 & 0.727 &    1.504 &       1.614 \\
QUINTA NORMAL       & 0.041 & 0.093 & 0.000 & 0.002 & 0.013 & 0.033 & 0.528 &    1.140 &       1.608 \\
RECOLETA            & 0.035 & 0.087 & 0.000 & 0.002 & 0.009 & 0.019 & 0.496 &    1.421 &       1.613 \\
RENCA               & 0.038 & 0.095 & 0.000 & 0.003 & 0.012 & 0.031 & 0.474 &    1.330 &       1.609 \\
SAN BERNARDO        & 0.034 & 0.096 & 0.000 & 0.001 & 0.005 & 0.019 & 0.559 &    1.824 &       1.602 \\
SAN JOAQU\'IN         & 0.038 & 0.087 & 0.000 & 0.005 & 0.015 & 0.035 & 0.510 &    
0.852 &       1.609 \\
SAN JOS\'E DE MAIPO   & 0.034 & 0.095 & 0.000 & 0.000 & 0.000 & 0.006 & 0.413 &    
0.152 &       1.616 \\
SAN MIGUEL          & 0.040 & 0.079 & 0.000 & 0.007 & 0.021 & 0.037 & 0.408 &    1.109 &       1.607 \\
SAN RAM\'ON           & 0.040 & 0.085 & 0.000 & 0.001 & 0.004 & 0.038 & 0.438 &    
1.007 &       1.605 \\
SANTIAGO            & 0.046 & 0.135 & 0.000 & 0.005 & 0.017 & 0.041 & 0.867 &    8.055 &       1.596 \\
VITACURA            & 0.026 & 0.047 & 0.000 & 0.002 & 0.007 & 0.021 & 0.210 &    0.423 &       1.616 \\
\~NU\~NOA               & 0.036 & 0.093 & 0.000 & 0.003 & 0.008 & 0.021 & 0.528 &    
1.949 &       1.609 \\
\hline
\end{tabular}
\end{table}

\begin{table}
\caption{The spectral radius as districts are sequentially added to a 
lockdown. Each value in the `Spectral radius' column corresponds to a lockdown
including all districts on or above the line where it appears in the
table. Spectral radii for only the first $9$ lockdowns have been shown in the
table.}
\label{tab:2}

\centering
\begin{tabular}{lr}
\hline
District  & Spectral radius \\
& after lockdown \\
\hline
PUENTE ALTO         & 1.559 \\
SANTIAGO            & 1.354 \\
EL BOSQUE           & 1.310 \\
PIRQUE              & 1.161 \\
MAIP\'U              & 1.111 \\
CERRO NAVIA         & 1.076 \\
LA CISTERNA         & 1.029 \\
\~NU\~NOA               & 0.984 \\
RECOLETA            & 0.946 \\
\multicolumn1c{$\vdots$} & \multicolumn1c{$\vdots$} \\
\hline
\end{tabular}
\end{table}

\begin{table}
\caption{The row sums (or local reproduction numbers $R_i$) for
districts after a lockdown is imposed on eight districts.}
\label{tab:3}

\centering
\begin{tabular}{ccc}
\begin{tabular}{lr}
\hline
District &     $R_i$ \\
\hline
CERRILLOS           & 0.611 \\
COLINA              & 0.549 \\
CONCHAL\'I            & 1.171 \\
ESTACI\'ON CENTRAL    & 0.892 \\
HUECHURABA          & 0.998 \\
INDEPENDENCIA       & 1.005 \\
LA FLORIDA          & 1.027 \\
LA GRANJA           & 1.144 \\
LA PINTANA          & 0.932 \\
LA REINA            & 0.695 \\
LAMPA               & 0.604 \\
LAS CONDES          & 0.810 \\
LO BARNECHEA        & 0.618 \\
LO ESPEJO           & 0.988 \\
LO PRADO            & 1.103 \\
MACUL               & 0.808 \\
\hline
\end{tabular}
&
\begin{tabular}{lr}
\hline
District &     $R_i$ \\
\hline
PADRE HURTADO       & 0.549 \\
PEDRO AGUIRRE CERDA & 0.963 \\
PE\~NAFLOR            & 0.535 \\
PE\~NALOL\'EN           & 0.863 \\
PROVIDENCIA         & 0.909 \\
PUDAHUEL            & 1.007 \\
QUILICURA           & 1.118 \\
QUINTA NORMAL       & 0.899 \\
RECOLETA            & 1.126 \\
RENCA               & 0.934 \\
SAN BERNARDO        & 0.879 \\
SAN JOAQU\'IN         & 0.863 \\
SAN JOS\'E DE MAIPO   & 0.825 \\
SAN MIGUEL          & 0.972 \\
SAN RAM\'ON           & 1.048 \\
VITACURA            & 0.731 \\
\hline
\end{tabular}
\end{tabular}
\end{table}


\newpage

\section*{Acknowledgments}

This work was supported by the Center for Mathematical Modeling ANID Basal
PIA program AFB 170001. JCM acknowledges support from FONDECYT INICIACION Grant
111170657 while MP and VA received support from the Center for Genome Regulation FONDAP 15090007. The authors would like to
thank Mauricio Canals of the School of Public Health in the Faculty of
Medicine at the University of Chile for many useful discussions concerning
epidemiological concepts.


\end{document}